\documentclass[aps,prb,reprint,amsmath,amssymb,superscriptaddress,floatfix]{revtex4-2}
\usepackage{xcolor}
\usepackage{graphicx}
\usepackage{dcolumn}
\usepackage{bm}
\usepackage[colorlinks, citecolor=blue, linkcolor=blue]{hyperref}

\usepackage{orcidlink}
\usepackage{txfonts}

\newcommand{\CNN}{Centre de Nanosciences et de Nanotechnologies, CNRS, Universit\'e Paris-Saclay, 91120 Palaiseau, France}
\newcommand{\LAF}{Laboratoire Albert Fert, CNRS, Thales, Universit\'e Paris-Saclay, 91767 Palaiseau, France}

\begin{document}

\title{Large frequency nonreciprocity of azimuthal spin-wave modes in submicron vortex state disks}

\author{Sali Salama\,\orcidlink{0009-0008-4274-0724}}
\email{sali.salama@universite-paris-saclay.fr}
\author{Joo-Von Kim\,\orcidlink{0000-0002-3849-649X}}
\affiliation{\CNN}
\author{Abdelmadjid Anane\,\orcidlink{0000-0001-5396-6165}}
\affiliation{\LAF}
\author{Jean-Paul Adam\,\orcidlink{0000-0003-2025-7105}}
\email{jean-paul.adam@universite-paris-saclay.fr}
\affiliation{\CNN}

\date{\today}

\begin{abstract}
Vortex states in thin film disks host spin-wave modes that are geometrically quantized according to their radial and azimuthal indices. Previous studies have shown that hybridization between these modes and the vortex core results in a sizable frequency nonreciprocity between low-order clockwise and counterclockwise propagating azimuthal modes. Here, we present a computational study of these spin-wave modes in submicron disks in which the spatial extension of the vortex core becomes comparable to the wavelength of certain modes. In such cases, we find that the frequency nonreciprocity can be large even for higher order radial and azimuthal indices, reaching several GHz and comparable to the mode frequencies themselves.
\end{abstract}

\maketitle

\section{Introduction}
Nonreciprocal propagation of spin waves is a common feature in ferromagnetic thin films, arising from the breaking of inversion symmetry. For instance, dipolar interactions govern magnetostatic surface spin waves (Damon-Eshbach modes)~\cite{damon1961magnetostatic}, which only propagate in a single direction at a specific surface. Similarly, chiral interactions of the Dzyaloshinskii-Moriya form result in a linear wave vector dependence in the dispersion relation, leading to frequency nonreciprocity, i.e., $f(\mathbf{k}) \neq f(\mathbf{-k})$~\cite{melcher1973linear, udvardi2009chiral, cortesortuno2013influence, moon2013spinwave}. This property has various applications in microwave engineering, including magnonic diodes, filters, circulators, and nonlinear magnonic resonators \cite{wang2020nonlinear, adam1990thin, grassi2020slow}.

Frequency nonreciprocity can also appear for spin waves around nonuniform magnetic textures, such as domain walls and vortices. These magnetic configurations have nontrivial topology but may not necessarily have a favored chirality. For instance, theoretical studies have shown that dipolar interactions within Bloch domain walls can lead to significant frequency nonreciprocity for spin waves confined along the wall center~\cite{henry2019unidirectional}. The nonreciprocal propagation there depends on the handedness of the propagation direction relative to the domains. In magnetic vortices, spin waves propagating along the azimuthal direction, either clockwise or counterclockwise around the core, are not degenerate in frequency for the same azimuthal index. In this case, the polarity of the vortex core, which determines whether the magnetic moments at the core center point upwards or downwards with respect to the film plane, determines the sign of this asymmetry. In contrast to the domain wall case, the mechanism for frequency asymmetry involving azimuthal spin-wave modes around vortex states has been a subject of debate, driven by several experimental~\cite{park2005interactions} and theoretical studies~\cite{ivanov1998magnon, guslienko2008dynamic}.

The cylindrical symmetry of the vortex state allows for the description of spin wave eigenmodes using two quantization mode numbers: the radial index $n$ and the azimuthal index $m$. These modes correspond to frequencies denoted by $f_{n,m}$. The frequency asymmetry is then defined as
\begin{equation}
    \Delta f_{n,m} \equiv f_{n,+m}-f_{n,-m}.
\end{equation}
Most studies have focused on the frequency asymmetry between the $(n,m)=(0,\pm 1)$ modes, $\Delta f_{0,1}$, since they couple strongly to the vortex core motion. Evidence for the pivotal role of the core was obtained in an experiment involving samples where the core was removed by etching away magnetic material from the central region of a thin film disk. This resulted in the suppression of the frequency asymmetry~\cite{hoffmann2007mode}. It was also reported that $\Delta f_{0,1}$ is approximately twice the gyrotropic mode frequency~\cite{zhu2005broadband}. However, models that assume interactions (exchange and dipolar) with a static core predict a smaller value of the asymmetry compared to the experiment~\cite{ivanov2005high}. Guslienko~\emph{et al.} \cite{guslienko2008dynamic} explained the asymmetry as a result of dynamic hybridization of the modes $n=0,\lvert m\lvert=1$ and the gyrotropic mode, which yields values close to the experiment~\cite{hoffmann2007mode}. Here, the hybridization refers to the spatial overlap in the mode profiles for the gyrotropic mode $\psi_G(r, \phi)$ and the azimuthal modes $\psi_R(r, \phi)$, which can be represented by an overlap integral of the form
\begin{equation}
N_v = \int_0^{2\pi} \int_0^R  \psi_G(r, \phi) \psi_R(r, \phi)\, r  \, dr \, d\phi,
\label{eq:overlapping_integral}
\end{equation}
where $(r,\phi)$ represent the spatial variables in cylindrical coordinates. Considering only the dependence on the azimuthal variable $\phi$, the spatial overlap is proportional to the quantity $\int_0^{2\pi} e^{\pm i\phi} e^{-im\phi} \, d\phi$, which vanishes for $|m| \neq 1$. Therefore, only spin-wave modes with $|m|=1$ are predicted to exhibit frequency nonreciprocity by this process~\cite{zhu2005broadband, park2005interactions}. The nonreciprocity has been found to depend linearly on the disk aspect ratio $\beta = d/R$, $\Delta f_{n,\pm 1} \propto \beta$ for $\beta < 0.1$, where $d$ is the film thickness and $R$ the disk radius. Awad~\emph{et al.} \cite{awad2010precise} generalized the nonreciprocity to all radial numbers $n$ and $|m|=1$ by assuming hybridization with the gyrotropic mode.

In this paper, we discuss results from micromagnetics modeling which demonstrate that the frequency nonreciprocity, $\Delta f_{n,m}$, is not limited to $|m| = 1$ modes but can be substantial for $|m| \neq 1$ in disks with small radii. In such disks, where the vortex core radius is a significant fraction of the disk radius, we find that $\Delta f_{n,m}$ exhibits a non-monotonous dependence on $n$, with a general increase in its magnitude. This phenomenon can be observed for various values of $m$. The underlying mechanism of this frequency nonreciprocity is the strong hybridization of the mode profiles with the vortex core, which tends to become more pronounced with larger radial indices.

\section{Geometry and method}
We consider a thin film disk in a vortex state with base values for the saturation magnetization $M_s$ of $140$ kA/m and exchange constant of $A$ of $3.7$ pJ/m, which are consistent with literature values for thin film yttrium iron garnet (YIG). The frequencies and spatial profiles of the spin-wave eigenmodes are computed using the open-source \texttt{magnum.np} code~\cite{bruckner2023magnum}, which uses a finite difference scheme to perform time integration of the Landau-Lifshitz equation. For the problem considered here, we use the built-in eigensolver routine which computes the first $N$ eigenmodes, ordered by frequency, by solving the eigenvalue problem of the linearized Landau-Lifshitz equation about the ground state. The disk studied has a nominal radius of $R=150$~nm and thickness of $d=65$~nm, which is discretized using 128$\times$128$\times$1 finite difference cells. Note that these are the base values for the geometric and micromagnetic parameters, as these parameters will be varied in the following in order to highlight their role in the frequency nonreciprocity.

\section{Results}
\begin{figure}
    \centering
    \includegraphics[width=1\linewidth]{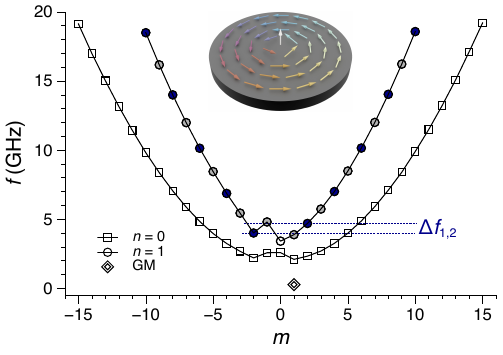}
        \caption{Vortex state eigenmode frequencies for a 150~nm radius disk as a function of the azimuthal number $m$ for two radial numbers $n=0,1$. The dark red diamond represents the gyrotropic mode (GM). Blue and gray dots represent different pairs of modes in the $n=1$ branch as a guide to the eye.  The inset illustrates the static magnetic configuration of the vortex state. The colors represent the in-plane magnetization distribution and the white upward arrow at the center denotes the polarity of the vortex. }
    \label{fig:disprel}
\end{figure}

A calculated dispersion relation for a 150~nm radius YIG disk in a vortex state with $p=1$ is shown in the inset of Fig.~\ref{fig:disprel}.
The two lowest frequency branches are shown, for $n=0$ and $1$, along with the singular point at $m = 1$ in the sub-GHz range that corresponds to the core gyration frequency. Here, positive azimuthal numbers refer to propagation in the counterclockwise direction around the disk when viewed from above, while negative $m$ describe clockwise propagation. The inset of the figure shows a schematic of the vortex state.

To better illustrate the frequency nonreciprocity between different $m$ modes for the $n=1$ branch in Fig.~\ref{fig:disprel}, we use the same color for identical $|m|$ and alternate between gray and blue for increasing $|m|$. $\Delta f_{n,m}$ in the figure indicates the asymmetry between the $(n,m) = (1,\pm 2)$ modes. By inspection, we can observe that the asymmetry has a different sign between the $(1,\pm 1)$ and $(1,\pm 2)$ modes, where $\Delta f_{1,1} < 0$ while $\Delta f_{1,2} > 0$. Nonreciprocity can also be discerned for higher-order azimuthal modes, further underscoring the fact that this phenomenon is not restricted to the well known $|m|=1$ case.

\subsection{Role of the disk size}
\begin{figure}
    \centering\includegraphics[width=8.5cm]{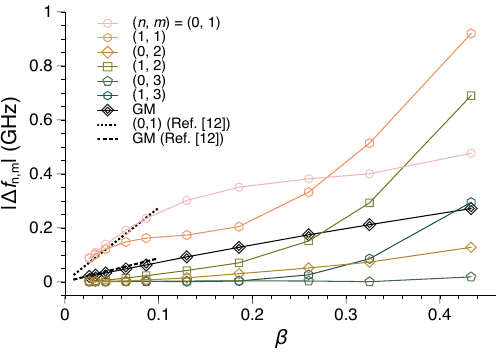}
    \caption{Frequency asymmetry, $\Delta f_{n,m}$, as a function of the aspect ratio, $\beta$, for different radial and azimuthal numbers, $(n,m)$. The film thickness is kept constant at $d=65$~nm. The black dotted line represents the predicted  $\Delta f_{0,1}$ from Ref.~\onlinecite{guslienko2008dynamic}. The black dashed represents the theoretical gyrotropic mode frequency according to Ref.~\onlinecite{guslienko2008dynamic}.}
        \label{fig:delta_f_vs_beta}
\end{figure}

To investigate the dependence of the frequency nonreciprocity on the disk size, we keep the thickness $d$ constant while varying the disk radius from 150~nm to 2.5~$\mu$m. The disks with radii ranging from 150~nm to 350~nm were discretized using 128×128×1 finite difference cells. For disks with radii between 500~nm and 1~$\mu$m, a discretization of 256×256×1 was used. While for disks with radii between 1.5~and 2.5~$\mu$m, the used discretization is 512$\times$512$\times$1. In Fig.~\ref{fig:delta_f_vs_beta}, we show $|\Delta f_{n,m}|$ for $n=0,1$ and $|m|=1,2,3$ plotted as a function of the aspect ratio $\beta=d/R$, along with the corresponding variation for the gyration frequency. 
Overall, we find that the magnitude of the frequency nonreciprocity increases with $\beta$. For a given $\beta$, it decreases as $m$ increases. For $\Delta f_{0,1}$, the initial variation for $\beta <0.1$ exhibits a linear behavior, which is consistent with the analytical calculations by Guslienko~\emph{et al.}~\cite{guslienko2008dynamic} as indicated by the red dashed line in Fig.~\ref{fig:delta_f_vs_beta}. Moreover, $\Delta f_{0,1}$ is larger than $\Delta f_{1,1}$ for $\beta<0.1$, in agreement with Ref.~\onlinecite{guslienko2008dynamic}.

For $\beta>0.1$, however, $\Delta f_{0,1}(\beta)$ exhibits a sub-linear dependence while $\Delta f_{1,1}(\beta)$ increases at a much faster rate, resulting in $\Delta f_{1,1}(\beta)>\Delta f_{0,1}(\beta)$ for $\beta>0.3$. We can also note that $\Delta f_{1,1}(\beta)$ increases monotonically and does not appear to reach a maximum within the range of aspect ratios considered, in contrast to the experimental results reported in Ref.~\onlinecite{awad2010precise}.

The gyration frequency is found to be linear with the aspect ratio in the range studied, consistent with analytical models. The dashed magenta line in Fig.~\ref{fig:delta_f_vs_beta} corresponds to the prediction by Guslienko \emph{et al.}~\cite{guslienko2002eigenfrequencies}, which agrees very well with the simulation results and could be extrapolated to account for the variation across the range of $\beta$ studied. The linear dependence of the gyration frequency on the aspect ratio stands in stark contrast to the variations observed for the different $\Delta f_{n,m}$, which suggests that while dynamical effects may play a role in the frequency nonreciprocity, it is not the dominant mechanism underpinning this asymmetry.

\subsection{Role of the exchange constant}
The size of the vortex core is governed by the exchange length, $\lambda_\mathrm{ex} = \sqrt{2A/(\mu_0 M_s^2)}$, where $A$ is the exchange constant and $M_s$ is the saturation magnetization. We can therefore examine how the size of the vortex core (relative to the disk size) influences the frequency nonreciprocity by varying either $A$ or $M_s$.

\begin{figure}
    \centering\includegraphics[width=1\linewidth]{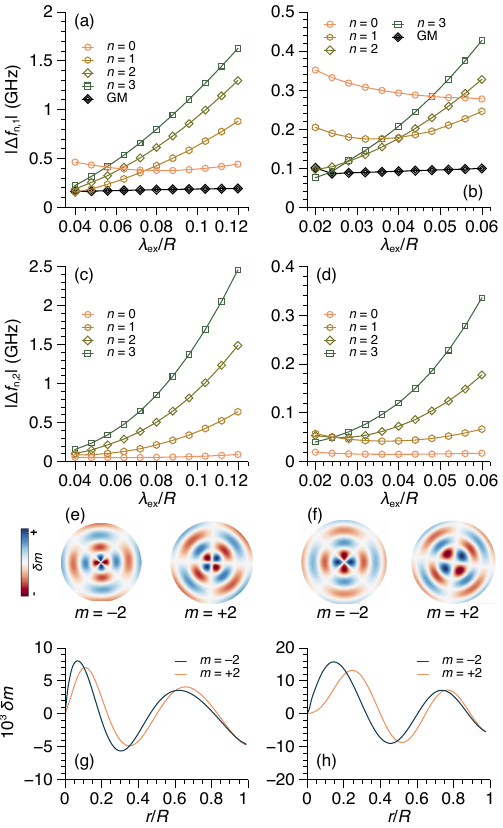}
    \caption{Frequency asymmetry as a function of the exchange length to disk radius ratio, $\lambda_\mathrm{ex}/R$ for two disk radii: (a) and (c) 250~nm, (b) and (d) 500~nm. The azimuthal mode is  $\lvert m\lvert$=1 in (a) and (b)  and $\lvert m\lvert$=2 in (c) and (d). $\lambda_\mathrm{ex}$ varies with $A$ at constant $M_s=140$~kA/m. The black curve GM in (a) and (b) represents the gyrotropic mode frequency. (e) and (f) show the spatial profiles of $n=3$, $\lvert m\lvert$=2 for $\lambda_{exch}/R = $ 0.02 and 0.12, respectively. The colors encode the amplitude of the dynamic magnetization perpendicular to the local equilibrium magnetization direction. (g) and (h) show the radial profiles of the modes in (e) and (f), respectively. }
        \label{fig:delta_f_vs_A}
\end{figure}

We first consider the role of the exchange constant by fixing the saturation magnetization at its nominal value of $M_s = 140$~kA/m. Figure~\ref{fig:delta_f_vs_A}(a) shows the frequency nonreciprocity for $|m|=1$ as a function of the ratio $\lambda_\mathrm{ex}/R$ by varying $A$ for a 250~nm radius disk which is discretized using 128$\times$128$\times$1 finite difference cells.
We observe two distinct behaviors. First, $\Delta f_{0,1}$ remains relatively constant across the range of $\lambda_\mathrm{ex}/R$ considered, exhibiting a small minimum around $\lambda_\mathrm{ex}/R \simeq 0.08$, while the nonreciprocity for $n>0$ increases monotonically.
For a larger disk radius of 500~nm which is discretized using 256$\times$256$\times$1 finite difference cells, however, both $\Delta f_{0,1}$ and $\Delta f_{1,1}$ exhibit a non-monotonic variation with the exchange length, while higher-order $n$ modes preserve the monotonic increase. Similar trends are also seen for the $m=2$ mode, where the variation of the nonreciprocity with the exchange length for the 250- and 500~nm radius disk is shown in Figs.~\ref{fig:delta_f_vs_A}(c) and \ref{fig:delta_f_vs_A}(d), respectively. Again, we can observe significant asymmetries in the GHz range, particularly for higher order $n$. The gyrotropic mode frequency remains relatively unaffected across the range of $\lambda_\mathrm{ex}/R$ considered as illustrated by the black curve in Figs.~\ref{fig:delta_f_vs_A}(a) and \ref{fig:delta_f_vs_A}(b).

For the 250~nm radius disk, the spatial profiles of the $(3,\pm 2)$ modes for two different values of $\lambda_{exch}/R$, 0.02 and 0.12, are depicted in Figures \ref{fig:delta_f_vs_A}(e) and \ref{fig:delta_f_vs_A}(f), respectively. In both cases, we can clearly discern a distinct difference in the mode amplitude near the disk center between the opposite azimuthal numbers. This difference is consistent with varying degrees of hybridization with the core. This phenomenon is particularly evident in Figures \ref{fig:delta_f_vs_A}(g) and \ref{fig:delta_f_vs_A}(h), where the mode profile is plotted as a function of the normalized radial distance from the disk center. The $m=-2$ mode, which propagates in the opposite direction to the core gyration, retains a significant amplitude near the disk center. Conversely, the first maximum for the $m=2$ mode exhibits a noticeable shift away from the center. This shift becomes more pronounced as $\lambda_\mathrm{ex}/R$ increases [Figure \ref{fig:delta_f_vs_A}(h)].

\subsection{Role of the saturation magnetization}
\begin{figure}
    \centering\includegraphics[width=1\linewidth]{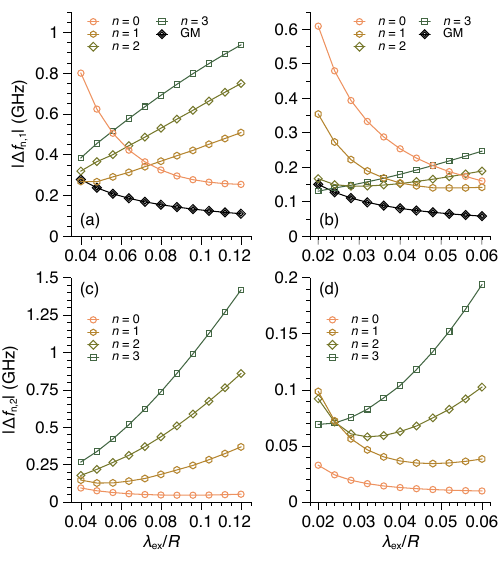}
    \caption{Frequency asymmetry as a function of the ratio of the exchange length to the disk radius ($\lambda_{exch}/R$) for two disk radii: (a) and (c) 250~nm, (b) and (d) 500~nm. The azimuthal mode is $\lvert m\lvert$=1 in (a) and (b) and $\lvert m\lvert$=2 in (c) and (d). $\lambda_\mathrm{ex}$ varies with $M_s$ at constant $A = 3.7$~pJ/m. The black curve ``GM'' in (a) and (b) represents the gyrotropic mode frequency.} 
                \label{fig:delta_f_vs_Ms}
\end{figure}

We now discuss the role of the dipolar field, which is captured by the magnitude of the saturation magnetization. We set the exchange constant to $A = 3.7$ pJ/m and consider the dependence of the frequency nonreciprocity on $\lambda_\mathrm{ex}$ by varying $M_s$. The results are presented in Fig.~\ref{fig:delta_f_vs_Ms}. 
In contrast to the previous case in Fig.~\ref{fig:delta_f_vs_A}, we can observe a strong decrease in $\Delta f_{0,1}$ as a function of the normalized exchange length for both 250- [Fig.~\ref{fig:delta_f_vs_Ms}(a)] and 500-nm [Fig.~\ref{fig:delta_f_vs_Ms}(b)] radius disks. A similar decrease is seen for $\Delta f_{1,1}$ in the 500-nm radius disk. This reduction is related to the diminishing dipolar interaction with increasing $\lambda_\mathrm{ex}$, which shows that it is the dominant mechanism for the hybridization for the low-order $m=\pm 1$ modes. Moreover, the gyrotropic mode is decreasing similarly as illustrated in Figs.~\ref{fig:delta_f_vs_Ms}(a) and \ref{fig:delta_f_vs_Ms}(b). For the higher-order $m=\pm 1$ modes, i.e., $n>0$ for the 250-nm radius disk and $n>1$ for the 500-nm radius disk, on the other hand, the nonreciprocity increases with $\lambda_\mathrm{ex}$, as seen previously in Figs.~\ref{fig:delta_f_vs_A}(a) and \ref{fig:delta_f_vs_A}(b). For these higher-order modes, the inversion symmetry breaking due to the dipolar interaction is amplified by the larger cores (with increasing $\lambda_\mathrm{ex}$), which results in similar asymmetries in the mode profiles for opposite $m$ as seen in Figs.~\ref{fig:delta_f_vs_A}(g) and \ref{fig:delta_f_vs_A}(h). These trends can be seen for the $m=2$ modes in Figs.~\ref{fig:delta_f_vs_Ms}(c) and \ref{fig:delta_f_vs_Ms}(d), where higher-order $n$ modes tend to increase with $\lambda_\mathrm{ex}$, with the non-monotonic behavior attributed to transitions between the dipolar-dominated and exchange-dominated regimes in agreement with previous results~\cite{uzunova2023nontrivial}.

\subsection{Nonreciprocity of higher-order modes}
\begin{figure}
    \centering\includegraphics[width=0.5\textwidth]{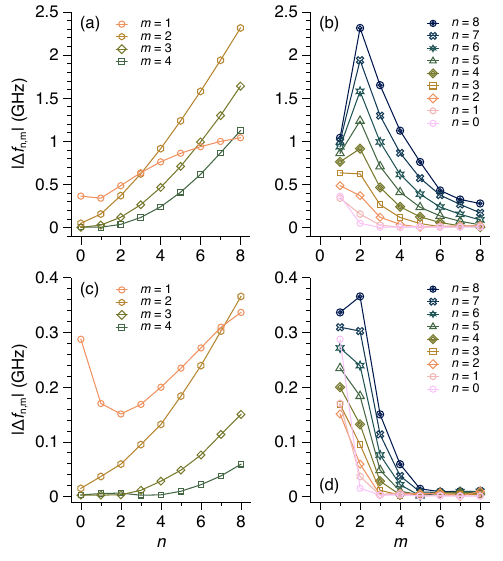} 
    \caption{Frequency asymmetry as a function of the radial number, $n$, for different azimuthal number, $m$, for disk radius of (a) 250~nm and (c) 500~nm. Frequency asymmetry as a function of $m$ for different $n$ for disk radius of (b) 250~nm and (d) 500~nm. Lines serve as guides to the eye.}
                \label{fig:delta_f_vs_nm}
\end{figure}

Having discussed how the exchange length influences the nonreciprocity, we return to the nominal system and examine how $f_{n,m}$ evolves with the two mode indices. The results are shown in Fig.~\ref{fig:delta_f_vs_nm} for a 250- and 500-nm radius disk. 
For the 250-nm radius disk, the nonreciprocity increases as a function of the radial index $n$ for the first four azimuthal indices, as depicted in Fig.~\ref{fig:delta_f_vs_nm}(a). The variation is mainly monotonous, with the exception for the $|m|=1$ modes for which a small minimum is seen at $n=1$. The evolution with the azimuthal index $m$ is presented in Fig.~\ref{fig:delta_f_vs_nm}(b). 
Overall, the nonreciprocity decreases for increasing $|m| \geq 2$ for the different radial indices considered. $|m| = 2$ is a notable transition point at which a maximum in the nonreciprocity is observed at large radial indices $n>3$.

Similar trends are seen for the 500-nm radius disk, as presented in Figs.~\ref{fig:delta_f_vs_nm}(c) and \ref{fig:delta_f_vs_nm}(d). For the variation with the radial index, we can observe a clearer minimum in the nonreciprocity for the $|m|=1$ modes, which occurs at $n=2$ here, unlike in the case of the 250-nm disk. For the dependence on the azimuthal index, a maximum in the nonreciprocity is also observed at $|m|=2$, but this time only for the largest radial index considered, $n=7$.

\section{Discussion and concluding remarks}
The ensemble of results reveals the widespread occurrence of frequency nonreciprocity across a wide range of radial and azimuthal numbers, which suggests that the argument based on the overlap integral in Eq.~\ref{eq:overlapping_integral} is insufficient. This deficiency stems from the fact that Eq.~\ref{eq:overlapping_integral}, which captures part of the energy related to the dynamical magnetization, assumes a scattering potential that is spatially-independent (with regard to the gyration and linear spin wave modes). In our case, however, the spatial extent of the core and the finite size of the disk cannot be neglected, which means that the energy acquires a strong spatial dependence. We can account for this with an overlap integral of the form
\begin{equation}
N_v = \int_0^{2\pi} \int_0^R  \psi_G(r, \phi) \psi_R(r, \phi) F(r, \phi) \, r \, dr \, d\phi,
\end{equation}
where $F(r,\phi)$ represents the scattering potential between the gyrotropic mode and the azimuthal spin waves. This potential can be expressed as a Fourier series, $ F(r, \phi) = \sum_{l} F_l(r) e^{i l \phi}$
where $l$ is the Fourier mode index of the scattering potential.
Expanding each term in azimuthal components, the integral over $\phi$ takes the form $\int_0^{2\pi} e^{i\phi (\pm 1 -l -m)} \, d\phi$, which is nonzero as long as $l + m = \pm 1$, thereby lifting the constraint of $|m| = 1$ for nonreciprocity.

The azimuthal dependence of the spatial overlap between the spin waves and the core can be understood as follows. The asymptotic behavior of the spatial profile of the magnons close to the vortex core is represented as $\sim r^{\mid q+m\mid }$ where $q$ is the chirality of the vortex~\cite{ivanov2002magnon}. Therefore, for large azimuthal numbers a node will be formed near the vortex core which decreases the interaction between the azimuthal spin-wave modes and the vortex core. Moreover, the spatial overlapping between the vortex core and the azimuthal modes increases by increasing the radial number which increases the corresponding nonreciprocity. Overall, these results serve to again highlight the fact that the nonreciprocity can be significant for higher-order radial and azimuthal mode indices, which suggests that mode hybridization with the vortex core affects a greater subset of spin-wave modes than previously anticipated.

Finally, frequency nonreciprocity is not unique to YIG but is a fundamental property of vortex states in confined structures. For instance, in a CoFeB disk with a radius of 100~nm, a thickness of 10~nm, $M_s=1250$~kA/m, and  $A=15$~pJ/m, the observed frequency nonreciprocity is even more pronounced than that of YIG. Specifically, the frequency nonreciprocity for the $n=6, |m|=2$ mode reaches 2.6 GHz. This suggests that the phenomenon should be observable experimentally over a wide range of material systems.
\\
\\
\begin{acknowledgments}
We thank T. Devolder and C. Chappert for fruitful discussions and comments on the paper. This work was supported by the Agence Nationale de Recherche under Contract Nos. ANR-10-LABX-0035 (Labex NanoSaclay, as part of the “Investissements d’Avenir” program) and ANR-20-CE24-0012 (MARIN).
\end{acknowledgments}
%


\end{document}